\documentclass[aps,prl,superscriptaddress,twocolumn,showpacs,preprintnumbers,amsmath,amssymb]{revtex4}
\usepackage[dvips]{graphicx,color}% Include figure files
\usepackage{dcolumn}% Align table columns on decimal point
\usepackage{bm}% bold math
\usepackage{ulem}
\usepackage{soul}
\begin{document}

\title{Extrinsic Spin Hall Effect Induced by Iridium Impurities in Copper}

\author{Y. Niimi}
\email{niimi@issp.u-tokyo.ac.jp}
\affiliation{Institute for Solid State Physics, University of Tokyo, 5-1-5 Kashiwa-no-ha, Kashiwa, Chiba 277-8581, Japan}
\author{M. Morota}
\affiliation{Institute for Solid State Physics, University of Tokyo, 5-1-5 Kashiwa-no-ha, Kashiwa, Chiba 277-8581, Japan}
\author{D. H. Wei}
\affiliation{Institute for Solid State Physics, University of Tokyo, 5-1-5 Kashiwa-no-ha, Kashiwa, Chiba 277-8581, Japan}
\author{C. Deranlot}
\affiliation{Unit\'{e} Mixte de Physique CNRS/Thales, 91767 Palaiseau France associ\'{e}e \`{a} l'Universit\'{e} de Paris-Sud, 91405 Orsay, France}
\author{M. Basletic}
\affiliation{Department of Physics, Faculty of Science, University of Zagreb, P.O. Box 331, HR-10002 Zagreb, Croatia}
\author{A. Hamzic}
\affiliation{Department of Physics, Faculty of Science, University of Zagreb, P.O. Box 331, HR-10002 Zagreb, Croatia}
\author{A. Fert}
\affiliation{Unit\'{e} Mixte de Physique CNRS/Thales, 91767 Palaiseau France associ\'{e}e \`{a} l'Universit\'{e} de Paris-Sud, 91405 Orsay, France}
\author{Y. Otani}
\affiliation{Institute for Solid State Physics, University of Tokyo, 5-1-5 Kashiwa-no-ha, Kashiwa, Chiba 277-8581, Japan}
\affiliation{RIKEN-ASI, 2-1 Hirosawa, Wako, Saitama 351-0198, Japan}

\date{March 22, 2011}

\begin{abstract}
We study the extrinsic spin Hall effect induced by Ir impurities in Cu 
by injecting a pure spin current into a CuIr wire 
from a lateral spin valve structure. 
While no spin Hall effect is observed without Ir impurity, 
the spin Hall resistivity of CuIr increases linearly 
with the impurity concentration. 
The spin Hall angle of CuIr, $(2.1 \pm 0.6)$\% throughout 
the concentration range between 1\% and 12\%, 
is practically independent of temperature. 
These results represent a clear example of predominant 
skew scattering extrinsic contribution to the spin Hall effect 
in a nonmagnetic alloy. 
\end{abstract}


\maketitle

The generation of pure spin currents, flows of only spin angular momentum 
without charge current, should play an important role in the next generation 
spintronic devices~\cite{maekawa}. 
The spin Hall effect (SHE) is one of the promising ways 
to create pure spin currents in nonmagnetic materials 
without using external magnetic fields or ferromagnets. 
The SHE was first predicted theoretically a long time ago~\cite{SHE1} and 
has recently received renewed interest 
which came from several theoretical predictions of SHE 
in nonmagnetic materials~\cite{SHE2,SHE3} 
and from the first experimental observation of the SHE 
in semiconductor systems using an optical method~\cite{kato_science_04}. 
By flowing the electric current into GaAs samples, spin-up and down electrons 
are accumulated on the opposite sides of the samples, 
which can be seen by scanning Kerr rotation microscopy. 
This is referred to as the direct spin Hall effect (DSHE). 
However, the spin Hall (SH) angle, which is defined as the ratio of 
the SH conductivity to the charge conductivity and represents 
the maximum yield of the transformation of charge into spin current density, 
is extremely small in semiconductors. Therefore an important challenge 
is to find more efficient materials for this transformation. 
Larger SHEs have been recently found in noble metals such as
Pt~\cite{saitoh_apl_06,kimura_prl_07,vila_prl_07,hoffmann_prl_10,morota} 
and Au~\cite{hoffmann_prl_10,seki_2008,hoffmann_prl_09} 
and this has triggered an important effort of research on the SHE 
in metallic materials. 

The SHE relies on spin-orbit (SO) interactions in materials and 
can be generated by intrinsic or extrinsic mechanisms. 
Recent theoretical works predict that the large SH angles 
of 4d and 5d transition metals, about 1\% in recent results on Pt 
for example~\cite{hoffmann_prl_10,morota}, stem from 
the intrinsic mechanism based on the degeneracy of d-orbits by SO 
coupling~\cite{kontani_jpsj_07,kontani_prb_08,kontani_prl_09}. 
This scenario has been supported by recent systematic experiments 
on the SHEs in 4d and 5d transition metals~\cite{morota}. 
The extrinsic SHE, on the other hand, relies on scattering by impurities 
(or other defects) presenting strong SO 
interactions~\cite{mertig,gu,fert_review}. 
There are two types of mechanisms, namely 
the skew scattering~\cite{skew_scattering} 
and the side jump~\cite{side_jump}. In the former case, 
the SH resistivity ($\rho_{\rm SHE}$) is proportional to the resistivity 
induced by the impurities ($\rho_{\rm imp}$), while, for side-jump effects, 
$\rho_{\rm SHE} \propto \rho_{\rm imp}^{2}$ when the impurities are 
the only source of resistivity or 
$\rho_{\rm SHE} \propto \rho_{\rm imp}\rho_{\rm total}$ 
when $\rho_{\rm total}$ includes an additional contribution 
from scattering potentials with weak SO interactions. A definite interest of 
the extrinsic SHE is that one can control the SH angle 
by changing the combination of host and impurity metals as well as 
by tuning the impurity concentration. In particular, 
the relation between the SHE and the resistivity can be studied 
not only by varying the temperature but also, in a much wider range, 
by changing the concentration of impurities. 

A series of pioneering works to this end had been performed in the 1980s 
by a part of the present authors using a ternary system consisting 
of a Cu matrix doped with a Mn spin polarizer and 5d impurities 
such as Lu, Ta, and Ir~\cite{fert_1981}. 
Large SH angles had been obtained, 
positive for CuIr (2.6\%) 
or negative for CuLu ($-1.2$\%), 
and had been ascribed to resonant scattering on 5d impurity states split 
into 5/2 and 3/2 levels by SO interaction. 
Therefore we put our focus on Ir as a strong SO scatterer. 
In order to determine the SH angle, either DSHE or 
inverse SHE (ISHE) is measured as follows: in DSHE experiments, 
the spins accumulated on the side surfaces of materials 
with strong SO interactions are detected with ferromagnetic contacts. 
In ISHE experiments, spin currents 
are converted into charge currents and then the potential drop along 
the current direction is detected. 
ISHE measurements have been intensively carried out 
in recent years by means of the pure spin current 
injection~\cite{kimura_prl_07,vila_prl_07,morota,seki_2008,hoffmann_prl_09,valenzuela_nature_06} or the microwave driven spin pumping 
techniques~\cite{saitoh_apl_06,hoffmann_prl_10}. 
In the present study we have adopted the spin absorption method 
using a lateral spin valve structure to measure the ISHE induced in Cu 
by Ir impurities. The final goal of the present study is 
to identify if the major contribution to the SHE is the skew scattering 
by the Ir impurities and what is the magnitude of the SH angle 
that can be obtained with such type of heavy impurity. 
We find that introducing Ir impurities to pure Cu 
which exhibits no SHE, increases the SH resistivity 
in proportion of the Ir concentration 
throughout the concentration range from 1\% to 12\%. 
This linear variation clearly shows that the skew scattering is the dominant 
mechanism in the CuIr alloys. The slope of a $\rho_{\rm SHE}$ vs 
$\rho_{\rm imp}$ plot gives the SH angle $\alpha_{\rm H}$ of 
$(2.1 \pm 0.6)$\% for CuIr.

\begin{figure}
\begin{center}
\includegraphics[width=7.5cm]{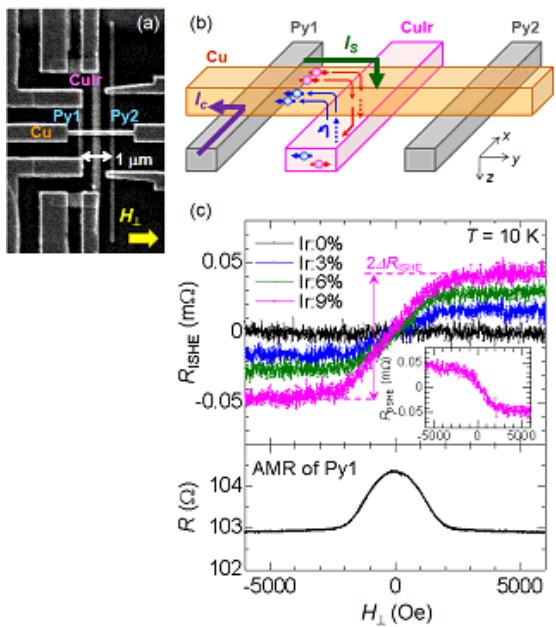}
\caption{(Color online) (a) Scanning electron microscopy image of a spin Hall device consisting of two Py wires and a CuIr middle wire bridged by a Cu wire. (b) Schematic of the mechanism of ISHE due to the spin absorption effect. (c) Inverse spin Hall resistance of CuIr with different Ir concentrations measured at $T=10$ K. For comparison, the direct spin Hall resistance of CuIr (9\%) is also shown in the inset. The bottom panel shows a typical AMR signal of Py1.} \label{fig1}
\end{center}
\end{figure}

Samples have been fabricated on a thermally oxidized silicon substrate 
using electron beam lithography on polymethyl-methacrylate resist 
and a subsequent lift-off process. 
We have used a lateral spin valve structure which consists of two Permalloy 
(Ni$_{81}$Fe$_{19}$; hereafter Py) wires (30 nm thick and 100 nm wide) and 
a CuIr middle wire (20 nm thick and 250 nm wide) bridged by a Cu wire 
(100 nm thick and 100 nm wide), as shown in Fig.~1(a). In this work, 
the distance between the two Py wires ($L$) is fixed to 1 $\mu$m and 
the CuIr wire is placed just in the middle of the two Py wires. 
To induce a difference between the switching fields of the two Py wires, 
one of them [Py1 in Fig.~1(a)] has two large pads at the edges. 
The Py wires were grown by electron beam evaporation, 
while the middle CuIr wires with different Ir concentrations 
(0\%, 1\%, 3\%, 6\%, 9\%, and 12\%) were deposited by magnetron sputtering. 
The Cu bridge was fabricated by a Joule heating evaporator 
using a 99.9999\% purity source. Prior to Cu evaporation 
a careful Ar ion beam etching (600 V beam voltage) was carried out 
for 1 min in order to clean the surfaces of Py and CuIr wires and 
to obtain highly transparent Ohmic contacts. 
Transport measurements were performed using a standard ac lock-in technique 
and a $^{4}$He flow cryostat. The magnetic field is applied along the hard 
and easy axes of Py for ISHE and nonlocal spin valve (NLSV) 
measurements, respectively. For each Ir concentration, 
at least three different samples from the same batch have been 
measured to check the reproducibility.

\begin{figure}
\begin{center}
\includegraphics[width=5.5cm]{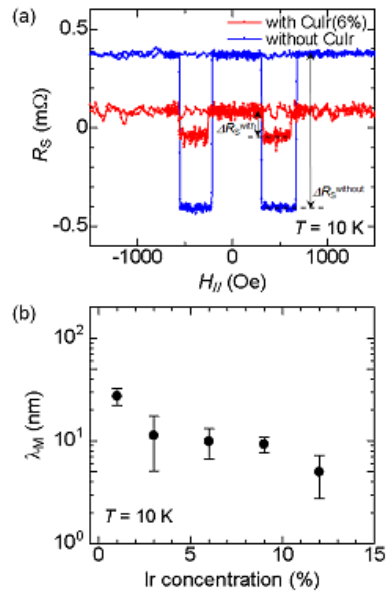}
\caption{(Color online) (a) NLSV signals measured at $T=10$ K with a CuIr (6\%) middle wire (red) and without CuIr wire (blue). (b) Spin diffusion length $\lambda_{\rm M}$ of CuIr at 10 K as a function of Ir concentration.} \label{fig2}
\end{center}
\end{figure}

First we discuss the ISHE results for CuIr with different Ir concentrations. 
The measurement circuit is depicted in Fig.~1(b). 
When the electric current flows from Py1 to the left side of the Cu wire, 
the resulting spin accumulation induces a pure spin current 
on the right side of the Cu wire. As we discuss in the next paragraph, 
a major part of the pure spin current is absorbed in the CuIr middle wire 
below the Cu wire since the spin diffusion length of CuIr ($\sim 10$ nm) 
is much smaller than that of Cu. 
The deflection in the same direction of the opposite 
spin-up and down vertical currents by skew scattering 
on the Ir impurities generates the ISHE signal. 
The ISHE resistance $R_{\rm ISHE}$ (equal to the ISHE voltage $V_{\rm ISHE}$ 
divided by the charge current $I_{\rm C}$), is plotted in Fig.~1(c) 
as a function of the magnetic field applied perpendicularly to the Py wires. 
$R_{\rm ISHE}$ increases linearly with the magnetic field up to $\sim$2000 Oe 
and then flattens off at the saturation of the magnetization of Py1  
%For higher fields the magnetization of Py1 is fully polarized 
[see the anisotropic magnetoresistance (AMR) curve of Py1 
in the bottom panel of Fig.~1(c)].
%, and $R_{\rm ISHE}$ saturates. 
It can also be seen in Fig.~1(c) that $R_{\rm ISHE}$ increases with 
increasing Ir. The inversion of the probe configuration 
[i.e., $I_{+} \Leftrightarrow V_{+}$, $I_{-} \Leftrightarrow V_{-}$ 
in Fig.~1(a)] enables one to measure the DSHE 
as previously reported~\cite{kimura_prl_07,vila_prl_07}. We could confirm 
that the SH resistance due to the DSHE is exactly the same as 
$R_{\rm ISHE}$ 
[see the inset of Fig.~1(c)]. 
This verifies the Onsager reciprocal relation in our system.

In order to estimate the spin diffusion length of CuIr and 
to use it in the evaluation of the spin current absorbed into the CuIr wire, 
we have measured the NLSV signal of our device. 
Note that in this case the magnetic field is applied along 
the easy axis of the two Py wires. 
As can be seen in Fig.~2(a), by inserting the CuIr middle wire, 
the spin accumulation signal $\Delta R_{\rm S}^{\rm with}$ 
($\equiv \Delta V_{\rm S}^{\rm with}/I_{\rm C}$) is reduced to 
$0.15\Delta R_{\rm S}^{\rm without}$ where $\Delta R_{\rm S}^{\rm without}$ 
is the spin accumulation signal without middle wires.
This indicates that most of the pure spin current 
injected from Py1 is absorbed in the CuIr wire. 
From the one-dimensional spin diffusion model~\cite{takahashi_prb_03}, 
the normalized spin signal 
$\Delta R_{\rm S}^{\rm with}/\Delta R_{\rm S}^{\rm without}$ 
can be expressed as follows; 
\begin{equation}
\frac{\Delta R_{\rm S}^{\rm with}}{\Delta R_{\rm S}^{\rm without}} 
\approx 
\frac{2R_{\rm M} \sinh (L/\lambda_{\rm N})}
{R_{\rm N} \left\{ \cosh (L/\lambda_{\rm N})-1 \right\} + 2R_{\rm M} \sinh(L/\lambda_{\rm N})} \label{eq1}
\end{equation}
where $R_{\rm N}$ and $R_{\rm M}$ are the spin resistances 
of Cu and CuIr middle wire, respectively.   
The spin resistance $R_{\rm X}$ of material ``X" is defined as 
$\rho_{\rm X} \lambda_{\rm X}/(1-p_{\rm X}^{2})A_{\rm X}$, 
where $\rho_{\rm X}$, $\lambda_{\rm X}$, $p_{\rm X}$ and $A_{\rm X}$ 
are respectively the electrical resistivity, the spin diffusion length, 
the spin polarization, and the effective cross sectional area 
involved in the equations of the one-dimensional spin diffusion 
model~\cite{note_polarization}. 
%$L$ is the distance between the two Py wires. As shown in Fig.~1(a), 
%$L$ is fixed to 1~$\mu$m. 
As reported previously~\cite{kimura_prl_08}, 
we can determine $\lambda_{\rm N}$, $\lambda_{\rm F}$, and $p_{\rm F}$ 
by measuring the NLSV signal without middle wire as a function of $L$. 
In the present study, 
$\lambda_{\rm N}=1.3$ $\mu$m, 
$\lambda_{\rm F}=5$ nm, and $p_{\rm F} = 0.23$ at $T=10$ K. 
Thus, we can extract the spin diffusion length $\lambda_{\rm M}$ 
of the CuIr middle wire from Eq.~(\ref{eq1}).
As can be seen in Fig.~2(b), $\lambda_{\rm M}$ drastically decreases 
with increasing the Ir atom.

We then calculate $\rho_{\rm SHE}$ 
as follows~\cite{maekawa,takahashi_review_08}:
\begin{equation}
\rho_{\rm SHE} = \frac{w_{\rm M}}{x} \left( \frac{I_{\rm C}}{\bar{I}_{\rm S}} \right) \Delta R_{\rm ISHE}
\label{eq2}
\end{equation}
where $\bar{I}_{\rm S}$ is the effective spin current injected 
(vertically for $\lambda_{\rm M} \ll w_{\rm N}$) into the CuIr wire and 
generating the ISHE, $w_{\rm M}$ is the width of CuIr wire and 
$x$ is a correction factor taking into account the fact 
that the horizontal current driven by the ISHE voltage 
balancing the SO deflections is partially shunted by the Cu wire 
above the CuIr/Cu interface. 
The correction factor $x$ is derived from additional measurements 
of the resistance of the CuIr wire with and without the interface with Cu 
and is found to be 0.36$\pm$0.08 for all the samples 
(see supplemental material~\cite{supplemental}); 
for the DSHE the same factor accounts for the shunting of 
the current through Cu.
$\Delta R_{\rm ISHE}$ is defined as the difference between $R_{\rm ISHE}$ 
at saturation field (above $\sim$2000 Oe) and $R_{\rm ISHE}$ at zero field
[see Fig.~1 (c)]. 
In our case $\lambda_{\rm M}$ is generally smaller 
than the thickness of the CuIr middle wire. 
The spin current injected from the interface with Cu 
decreases in the CuIr wire, 
exponentially in the limit $\lambda_{\rm M} \ll t_{\rm M} \ll w_{\rm N}$, 
linearly down to zero at the bottom of CuIr for 
$t_{\rm M} \ll \lambda_{\rm M} \ll w_{\rm N}$, 
the general expression of $\bar{I_{\rm S}}/I_{\rm C}$
for values of $t_{\rm M}$ (20 nm) and $\lambda_{\rm M}$ 
[$5-27$ nm in Fig.~2(b)] much smaller than $w_{\rm N}$ (100 nm) 
being~\cite{morota};
\begin{widetext}
\begin{eqnarray}
\frac{\bar{I_{\rm S}}}{I_{\rm C}} &\equiv& 
\frac{\int_{0}^{t_{\rm M}} I_{\rm S}(z) dz}{t_{\rm M} I_{\rm C}} 
=\frac{\lambda_{\rm M}}{t_{\rm M}}\frac{\left( 1-e^{-t_{\rm M}/\lambda_{\rm M}} \right)^{2}}{1-e^{-2t_{\rm M}/\lambda_{\rm M}}} \frac{I_{\rm S}(z=0)}{I_{\rm C}} \nonumber\\
&\approx& \frac{\lambda_{\rm M}}{t_{\rm M}}\frac{\left( 1-e^{-t_{\rm M}/\lambda_{\rm M}} \right)^{2}}{1-e^{-2t_{\rm M}/\lambda_{\rm M}}} \frac{2 p_{\rm F} R_{\rm F} \sinh \left( L/2\lambda_{\rm N} \right) } { \left[ R_{\rm N} \left\{  \cosh \left( L/\lambda_{\rm N} \right) - 1 \right\} + 2 R_{\rm F} \left( e^{L/\lambda_{\rm N}}-1 \right) \right] +2 R_{\rm M} \sinh \left( L/\lambda_{\rm N} \right)}. \label{eq3}
\end{eqnarray}
\end{widetext}
By using Eqs.~(\ref{eq2}) and~(\ref{eq3}) we can derive the SH resistivity 
$\rho_{\rm SHE}$ from $\Delta R_{\rm ISHE}$. In Fig.~3 
we plot $\rho_{\rm SHE}$ of CuIr as a function of the resistivity 
induced by the Ir impurities, i.e. $\rho_{\rm CuIr}-\rho_{\rm Cu}$. 
It nicely follows a simple linear dependence up to Ir concentration of 12\%. 
This clearly shows that the dominant mechanism of the extrinsic SHE 
induced by the Ir impurities is the skew scattering. 
The SH angle characteristic of this skew scattering, 
$\alpha_{\rm H} = \rho_{\rm SHE}/\rho_{\rm imp}$, is $(2.1 \pm 0.6)$\%. 
In the previous measurements of the ISHE due to the skew scattering 
induced by Ir impurities in Cu after spin-polarization of the current 
by dilute Mn impurities 
(Mn impurities alone not contributing to the Hall effect), $\alpha_{\rm H}$ 
was 2.6\%, which is quantitatively consistent with 
our result~\cite{fert_1981}. 
%Actually our experimental technique can lead to 
%some underestimation of the SHE. In DSHE measurements some fraction of 
%the current in CuIr flows in Cu above the CuIr/Cu interface 
%so that the current effectively contributing to the SHE below 
%the CuIr/Cu interface is smaller than the current $I_{\rm C}$ flowing in CuIr 
%out of the contact region. In ISHE experiments a similar effect occurs: 
%the ISHE voltage needed to balance the deflection of the injected 
%vertical spin current is smaller than it would be without the additional 
%conduction in Cu. 
%We can take into account this current reduction experimentally 
%by comparing the voltage drops along the CuIr wire with and without 
%the shunting Cu. The obtained current reduction is about 0.36 
%almost independent of the Ir concentration. This consideration 
%results in the re-scaled SH angle of 2.1\% for our measurements. 

\begin{figure}
\begin{center}
\includegraphics[width=7cm]{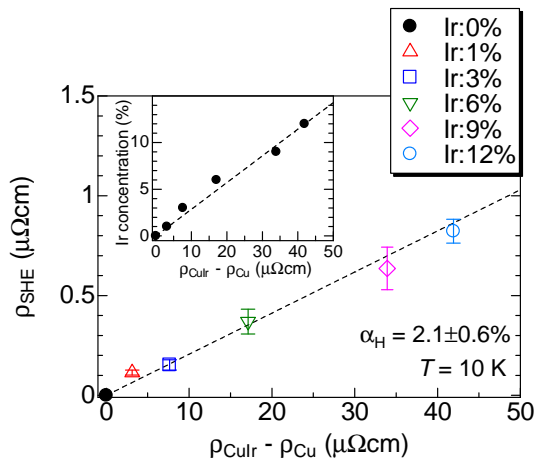}
\caption{(Color online) Spin Hall resistivity $\rho_{\rm SHE}$ as a function of the resistivity induced by Ir impurities, i.e., $\rho_{\rm CuIr}-\rho_{\rm Cu}$ at $T=10$ K. The error bar along the $y$ axis is found by calculating the standard deviation among at least three different samples on the same batch. The error bar for $x$ axis is within the dot. The inset shows $\rho_{\rm CuIr}-\rho_{\rm Cu}$ vs Ir concentration in Cu.} \label{fig3}
\end{center}
\end{figure}

\begin{figure}
\begin{center}
\includegraphics[width=5.5cm]{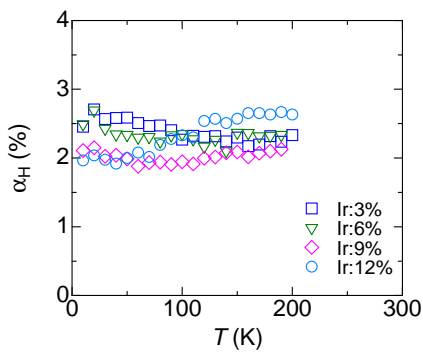}
\caption{(Color online) Temperature dependence of spin Hall angle of CuIr with different Ir concentrations.} \label{fig4}
\end{center}
\end{figure}

%We have also measured the temperature dependence of $\alpha_{\rm H}$. 
As shown in Fig.~4, the SH angle changes only weakly as a function of 
temperature. This is an additional proof for the mechanism of skew scattering 
by impurities since the contributions from intrinsic SHE or 
impurity scattering with side jump would be affected by the temperature 
dependence of the total resistivity. 
Finally, let us mention some results we obtained on the SH resistivity 
of AgIr. 
%We have performed a similar measurement by replacing Cu with Ag. 
In this case, the estimated SH angle is definitely smaller, 0.6\%. 
This large reduction is probably due to the very small solubility of Ir 
in Ag~\cite{karakaya}. 

In conclusion, we have measured the SH resistivity of the SHE induced 
by Ir impurities in Cu. The SH resistivity $\rho_{\rm SHE}$ is approximately 
proportional to the impurity-induced resistivity $\rho_{\rm imp}$ and 
practically temperature independent, which allows us to ascribe it 
to skew scattering on the Ir impurities. For the SH angle, 
characteristic parameter of the transformation of charge into spin current, 
we find 2.1\%, which is quantitatively consistent with 
the value derived in previous experiments on CuIr, 2.6\%~\cite{fert_1981}. 
%However we have presented arguments on the probable 
%underestimation of the SHE 
%with our experimental technique.
%Taking into account the shunting effect due to the Cu wire results 
%in the re-scaled SH angle of 2.1\% for our measurements.
Such values of the SH angle are larger than 
those obtained with pure metals~\cite{saitoh_apl_06,kimura_prl_07,vila_prl_07,hoffmann_prl_10,morota} and confirm that scattering 
by impurities is a very promising way to obtain large SH angles, 
as it is predicted by several recent skew scattering calculations 
papers~\cite{mertig,gu,fert_review}. 
Fert and Levy~\cite{fert_review} have calculated the contributions 
from both skew scattering and scattering with side jump on impurities in Cu.
For Ir impurities, they predict predominant skew scattering effects 
in the concentration range of our experiments, in agreement with our results. 
However, for other types of impurities (Os, Ta) in Cu, they find that the 
side-jump contribution to the SH angle can be definitely larger, 
that is a few percent for concentrations in the 1\% range 
and therefore above 10\% 
for concentrations in the 10\% range~\cite{fert_review}. 
Alloys combining side-jump and skew scattering effects 
in such a concentration range are promising to obtain a large SH angle 
and an efficient transformation of charge current into 
spin current in devices without magnetic components.

We acknowledge helpful discussions with H. Jaffres, S. Takahashi, and 
S. Maekawa, and numerical simulations 
performed by P. Metaxas and P. Bortoloti. 
We would also like to thank Y. Iye and S. Katsumoto 
for the use of the lithography facilities. 
This work was supported by KAKENHI %(Grant No. 22840012) 
and a Grant-in-Aid for Scientific Research in Priority Area 
%``Creationand Control of Spin Current" 
%(Grant No. 19048013)
from MEXT.
%the Ministry of Education, Culture, Sports, Science and Technology of Japan.


\begin{thebibliography}{00}
\bibitem{note_polarization}
$p_{\rm N}$ and $p_{\rm M}$ are zero since Cu and CuIr are nonmagnetic metals. $A_{\rm N} = w_{\rm N}t_{\rm N}$ for $\lambda_{\rm N} \gg t_{\rm N}$, $A_{\rm F} = w_{\rm F}w_{\rm N}$ for $\lambda_{\rm F} \ll t_{\rm F}$, and $A_{\rm M} =w_{\rm M}w_{\rm N} \tanh(t_{\rm M}/\lambda_{\rm M})$ for $t_{\rm M}, \lambda_{\rm M} \ll w_{\rm N}$. The product $A_{\rm N}\lambda_{\rm N}$ ($A_{\rm F}\lambda_{\rm F}$, $A_{\rm M}\lambda_{\rm M}$) characterizes the effective volume of spin relaxation in N (F, M).
\end{thebibliography}
\end{document}